\documentclass[prb,preprint,12pt, preprintnumbers]{revtex4-1}

\usepackage{amsmath}  
\usepackage{amsfonts} 
\usepackage{graphicx} 

\usepackage{epstopdf}

\begin{document}

\preprint{CALT-TH-2014-143}
\preprint{HDP: 14 -- 04}

\title{The plucked string: an example of non-normal dynamics}

\author{David Politzer}
\email{politzer@theory.caltech.edu}
\affiliation{California Institute of Technology, 452-48 Caltech, Pasadena, CA 91125}

\date{August 20, 2014}

\begin{abstract}
Motion of a single Fourier mode of the plucked string is an example of  transient, free decay of linear, coupled, damped oscillators.  It shares the rarely discussed features of the generic case, e.g., possessing a complete set of non-orthogonal eigenvectors and no normal modes, but it can be analyzed and solved analytically by hand in an approximation that is appropriate to musical instruments' plucked strings.
\end{abstract}

\maketitle

\section{Background}

The transient, free decay of coupled, damped oscillators is not discussed in elementary physics courses and rarely, if ever, in advanced ones.  The discussion in advanced physics textbooks is cursory, typically suggesting that one would proceed ``just as one might imagine" but that the details are cumbersome.  The new features possessed by the generic or typical version of such systems relative to the well-studied ones are just not part of the basic physics education offered to all students of physical sciences and engineering.  ``Proportional" damping (where each normal mode has its own individual damping constant) is often presented and is easy to analyze, but it is a very special subset of possible damping in coupled systems, and it misses some of the most dramatic features that are actually quite common.  With no  basic understanding provided by the elementary physics canon, when analogous aspects arise in particular situations, the people involved sense an aspect of discovery.  Sometimes the ``newly" discovered perspective  has had major impact.  Although the mechanics of such linear systems has been understood, in principle, for hundreds of years, rediscovery of their special features has occurred even into the $21^{\text{st}}$ Century.

Here are some examples of such rediscoveries.  Understanding the $K^o$--$\bar K^o$ meson system\cite{MGM} in the 1950's laid the groundwork for the experiments that identified CP and T (time reversal symmetry) violations in the fundamental interactions.\cite{CP}$^,$\cite{kabir}  Mechanical engineers in the 1960's were interested in earthquake shaking of buildings,\cite{caughey}$^,$\cite{caughey2} which can exhibit something now known as transient growth after the initial quake ground shaking has ceased.  The origin of the characteristic sound of the piano\cite{weinreich} (in contrast to earlier stringed, keyboard instruments) was elucidated in the 1970's.  Stability analyses in fluid mechanics and analogous problems in applied linear algebra witnessed a major revolution starting in the late 1990's.\cite{schmidt}$^,$\cite{tref}  ``Transient growth" is the generic term applied wherever all eigenvalues and eigenvectors decay exponentially in time and yet some combinations initially grow before ultimately dying off.  And additional examples of systems and situations of this type continue to be identified.

Of course, numerical integration of differential equations has gotten easier and better over the years.  And it has often been observed that what are now identified as ``non-normal" linear systems sometimes exhibit surprising transient behavior, quite sensitive to parameters and initial conditions.  But that is not a substitute for thorough understanding of at least one simple, mechanical system.

What is needed is an explicit example that exhibits the phenomena particular to the {\it simultaneous} presence of {\it both} coupling and damping and yet is visualizable and analytically tractable.  With respect to coupling, as simple as possible means only two oscillators. With respect to damping, as simple as possible means weak damping.  But solution of the general problem of this form involves solving a quartic polynomial.  Quartic is the highest order polynomial for which a closed-form solution exists.  That solution has been known since the $16^{\text{th}}$ Century, but it is far longer than most people can remember or comprehend at a glance.  Reducing the algebra to quadratic equations requires weak coupling as well.  But even that is not enough.  We will need a system whose frequencies, before coupling and damping, are degenerate or nearly degenerate, i.e., the same.  A bonus of having the coupling and damping be weak is that the combined behavior shows vestiges of the familiar separate problems of the single damped oscillator and of undamped coupled oscillators.

\begin{figure}[h!]
\centering
\includegraphics[width=2.8in]{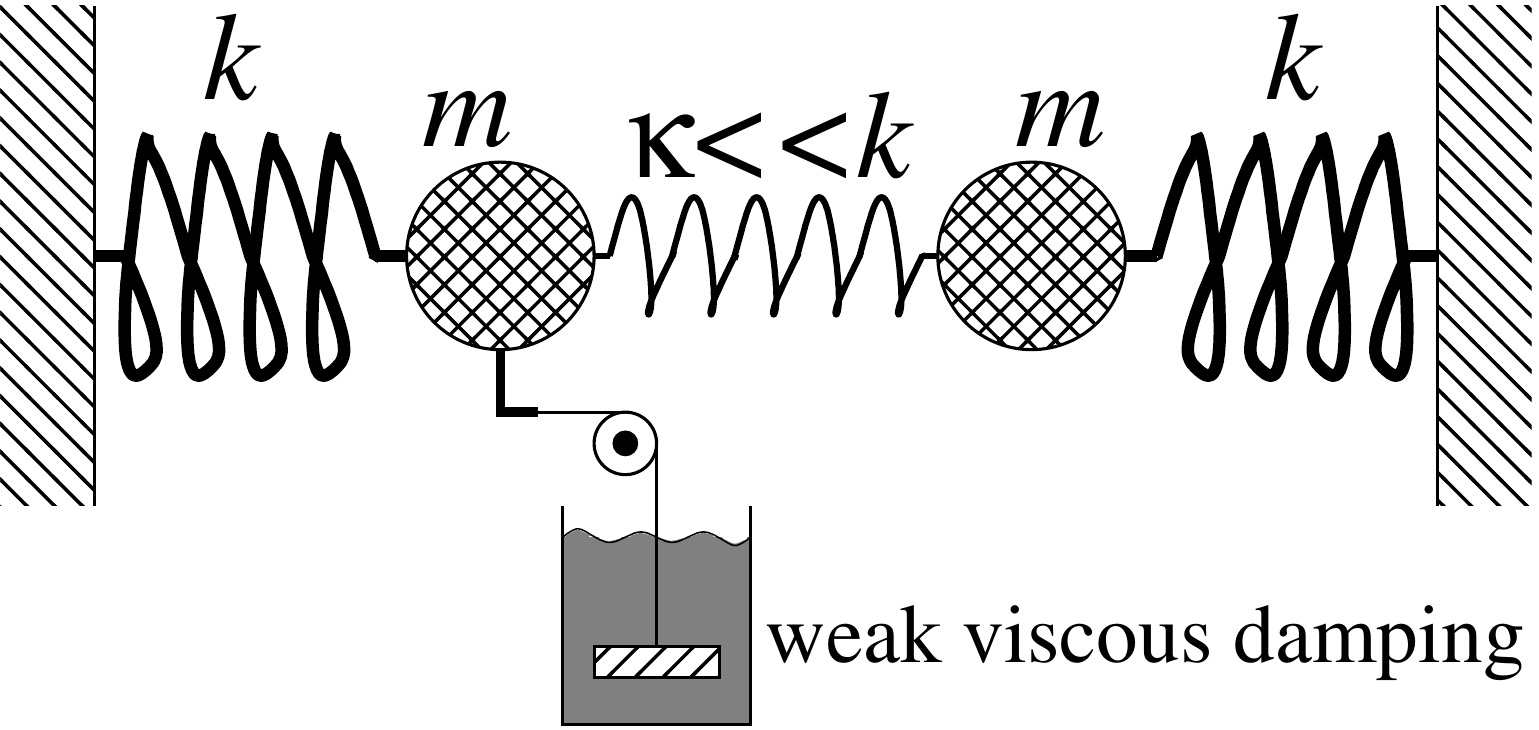}
\caption{two oscillator mechanical analog}
\label{springs}
\end{figure}

A mechanical system satisfying those requirements is illustrated in Fig.~\ref{springs}.
\noindent Two identical oscillators are coupled with a weak spring, $\kappa$, but only one of the oscillators is weakly damped.  With sufficient care, such an apparatus could be constructed.  In fact, Bruce Winstein\cite{winstein}, when he gave colloquia and talks about CP violation, brought along just such a mechanical model that he had made, which included finely adjustable damping and coupling.  Ref.~\onlinecite{quant} makes explicit the connection of the neutral kaons to a two coupled degenerate mode damped system.

On the other hand, plucking a typical stringed instrument readily provides the appropriate degrees of freedom and parameter ranges.

As background, it is useful to understand the generic case of two weakly coupled, degenerate oscillators.  In typical demonstrations, damping is minimized.  These could be springs,\cite{springs} pendulums,\cite{pendulums} or, most charmingly, the Wilburforce pendulum.\cite{wilbur}  Essential features are the small splitting of the normal frequencies, identification of the normal modes, and dependence of the qualitative motion on initial conditions --- in particular, the possibility of beats.

The other essential bit of background is the decomposition of string motion into normal modes.  The ideal string can be analyzed as a set of normal modes, which are uncoupled oscillators whose frequencies are integer multiples of the lowest, ``fundamental" frequency.  What is often not mentioned is that for a physical string, each of these modes is actually a degenerate pair, corresponding to the possible vertical and horizontal motion at the given frequency.  In the simplest approximation, these are totally independent motions, which are then superposed into the motion of the single physical string.

And for the string actually to be ``musical," the damping and the coupling must be {\it both} small {\it yet} non-negligible.

There are several related meanings of the term ``non-normal."  Here, it refers simply to matrices with complete sets of eigenvectors which are not all orthogonal to one another.  They are common elements of descriptions of a variety of systems including ones that exhibit transient growth where all eigenvectors, individually, decay monotonically.\cite{schmidt}$^,$\cite{tref}$^,$\cite{rahul}  As described in the following, the total energy of two coupled, damped oscillators decays in time, no matter what the initial conditions or parameter values.  However, there are ranges of parameters and initial conditions for which the amplitude and energy of {\it one} of the oscillators can grow before eventually decaying.  In such cases, the decay of the total system's energy can come in spurts, rather than a steady one- or two-rate exponential.

There are, of course, quantum mechanical analogs, where more sophisticated methods and different approximations yield a complementary set of solutions.\cite{quantum}

\section{The String as Two Coupled, Damped Oscillators}

The ideal string stretched taut between two fixed points has normal modes with frequencies that are integer multiples of the fundamental.  In a musical instrument, there are actually two degenerate modes for each frequency, reflecting the possibility of string displacements in the plane transverse to the string direction.

There are always some further interactions that couple those degenerate modes.  The result is that their frequencies are split an amount proportional to the strength of that coupling.  The coupling has the further consequence of establishing the form of the actual normal modes, i.e., the motions that have precisely just one of those two frequencies.  And the modes thus determined  are not coupled.  (It is said that the perturbation diagonalizes the interaction.)  In the specific context of a stringed instrument, it is important that the splitting be small so that any combination of the two is perceived as essentially a single pitch.

By itself, the vibration of a string produces almost no sound --- because the tiny cross section of the string moves almost no air.  There must be some further transduction of the string motion to air motion.  In acoustic instruments, that is accomplished by linking one end of the string to a sound board.  String oscillations force sound board motion, which in turn produces sound.  Hence, at least one ``fixed" end of the string is not actually fixed.  Furthermore, in a good musical instrument, that end motion is the string's primary loss of energy.  And again, that damping must be weak so that the consequent width or spread in frequency due to the damping leaves a single discernible pitch rather than noise.

So, for the present, we focus on a single original frequency.  The system is approximated as two initially degenerate oscillators which are weakly coupled and weakly damped.  The catch is that the coupling and the damping, represented as matrices in the configuration space of the two initial oscillators, are generally not simultaneously diagonalizable.  If the system is a particular mode of a string, we may choose as a basis the up-and-down motion (relative to the sound board) and the side-to-side.  

The coupling of vibration to the sound board is typically far more effective for vertical motion than for horizontal.  To make matters as simple as possible, we will consider explicitly small vertical damping (denoted by $\gamma$) and no horizontal damping.  If the system is started out in one of the modes of the undamped, coupled problem, then the non-zero damping will decrease the amount of vertical motion relative to the horizontal.  Unless the undamped modes were precisely pure vertical motion for one mode and pure horizontal motion for the other, this damping would disturb the balance that defined the normal mode.  Thus, the damping mixes in some amount of the other  mode.

What is it that actually happens?

We choose the original restoring forces, the coupling, and the damping so that the system is linear (e.g., with the damping proportional to the velocities).  With positive damping, the total energy must decrease monotonically with time.  It is handy (and virtually essential) to use a complex number representation of the frequencies and displacements; their superposition into real motions at the end is totally parallel to standard treatments of the free decay of the single damped harmonic oscillator.

We will find that there are, indeed, two eigenfrequencies.  They describe exponential decay multiplied by sinusoidal oscillation.  They have corresponding eigenvectors, which are possible motions that follow their single eigenfrequency.  However, these vectors have complex components, which, translated into the real motion of  strings, means that their motion in the transverse plane is elliptical.  And, finally, these eigenvectors are not orthogonal.  One consequence is that the total energy and the rate of energy dissipation (which is the volume of produced sound in this simple model of a string) are not necessarily the sum of two independent, decreasing exponentials.

Newton's second law yields two coupled differential equations in time that are linear in the two displacements.  Write the displacements, $x_1(t)$ and $x_2(t)$ in vector form with 
\begin{equation}
\nonumber
	\mathbf{x}(t)
	=
\begin{bmatrix}
	x_1(t)\\
	x_2(t)\\
\end{bmatrix}.
\end{equation}
\noindent For the plucked string, $x_1(t)$ is a given mode's vertical motion, and $x_2(t)$ is the horizontal motion.

The equations of motion take the form
\begin{equation}
\label{eq-mot}
	\mathbf{\ddot x} = - \mathbf{K} \cdot \mathbf{x} - \mathbf{\Gamma} \cdot \mathbf{\dot x} ~  ,
\end{equation}
where {\bf K} and  $\mathbf{\Gamma}$ are $2\times2$ matrices representing the zeroth order restoring forces, the coupling, and the damping.  (The general mathematical problem would include a mass matrix multiplying $\mathbf{\ddot x}$.)

The unit of time can be chosen to make the zeroth order {\bf K$_\text{o}$  =  I}, the unit matrix.  The most general possible form of {\bf K} is Hermitian, which means it has three real parameters for this mode pair.  (The fourth parameter is absorbed into the average frequency, which is normalized to 1.)  For the present, however, choose just the simplest possible mathematical form the has the feature that the eigenvectors of {\bf K} are mixed by the damping.  That simplest, one parameter form is
\begin{equation}
\nonumber
	\mathbf{K}
	=
	\begin{pmatrix}
	1 & ~ ~ \epsilon \\
	\epsilon & ~ ~ 1
	\end{pmatrix} ~.
\end{equation}
This particular {\bf K} describes the mass/spring system of Fig.~\ref{springs}.  From the standpoint of actual stringed instruments, it is theoretically possible but may be an oversimplification.  However, with only a single parameter in {\bf K}, the ensuing math is far easier to follow.  At the end, it will also be easy to see how the same calculations with the most general possible {\bf K} would go through and see what aspects of the results are generic.

The damping, as described above, takes the form
\begin{equation}
\nonumber
	\mathbf{\Gamma}
	=
	\begin{pmatrix}
	\gamma & ~ ~ 0 \\
	0 & ~ ~ 0
	\end{pmatrix} ~ .
\end{equation}
For weak coupling and damping, units are chosen such that all of the angular frequencies are close to 1.
There is no basis in which {\bf K} and $\mathbf{\Gamma}$ are both diagonal.  A commonly recognized reflection of this is that the commutator
\begin{equation}
\nonumber
	[\mathbf{K}, \mathbf{\Gamma}] = \mathbf{K} \cdot \mathbf{\Gamma} - \mathbf{\Gamma} \cdot \mathbf{K} =
	\begin{pmatrix}
	0 & ~ ~ - \epsilon \gamma \\
	\epsilon \gamma & ~ ~ 0
	\end{pmatrix}
	\ne 0 ~ .
\end{equation}	
(Of course, there are other forms of damping for which the damping and coupling matrices do commute and are simultaneously diagonalizable.  This is known as proportional damping, and the separation into normal modes is straightforward.)

\section{Solution: Eigenvalues}

We seek eigenvalues $\alpha$ and time-independent eigenvectors $\mathbf{x}_o$ such that e$^{\alpha t} \mathbf{x}_o$ is a solution to Eq.~(\ref{eq-mot}).  Plugging that in yields
\begin{equation}
\label{plug-in}
	(\alpha^2 \mathbf I + \mathbf K + \alpha \mathbf \Gamma) \cdot \text{e}^{\alpha t} \mathbf{x}_o = 0 ~,
\end{equation}
where {\bf I} is the identity matrix.
$\mathbf{x}_o = 0$ is a solution to Eq.~(\ref{plug-in}) but not to the problem at hand.  For all other solutions, the matrix factor in Eq.~(\ref{plug-in}) cannot have an inverse, and that requires that its determinant vanish, which is the following ``characteristic equation":
\begin{equation}
\label{det}
	(\alpha^2 + 1)^2 + \gamma \alpha (\alpha^2 + 1) - \epsilon^2 = 0 ~.
\end{equation}
  
This would be easy to solve were either $\gamma$ or $\epsilon$ zero, but as it stands it is a quartic equation.  Its analytic solution is hundreds of characters long and contains many nested square and cube roots.  Most people find it impossible to discern by inspection the qualitative properties for particular values of the parameters.  Just as the single damped oscillator has cases with radically different qualitative behavior, i.e., over damped, under damped, and critically damped, there are cases here, too --- only many more.

For weak coupling and weak damping, we can anticipate the structure of the solutions from physical considerations.  With $0 < \gamma \ll 1$, the four solutions for $\alpha$ will be two complex conjugate pairs.  Re[$\alpha$] will be negative, reflecting the monotonic loss of energy.  Im[$\alpha$] comes in conjugate pairs.  These conjugate pair solutions can ultimately be superposed to get real solutions with sines and cosines of $t$ with the same frequency.  And this multiplicity of solutions allows fitting of any initial conditions of the two oscillators.

If {\it both} $\gamma \ll 1$ and $\epsilon \ll 1$, then {\it all} the frequencies will be near to 1, i.e. {\it i} Im[$\alpha] \approx \pm i$.  This offers a way to approximate Eq.~(\ref{det}) and reduce the algebra problem to a quadratic.\cite{weinreich2}  In the term $\gamma \alpha (\alpha^2 + 1)$, approximating the first $\alpha$ by $\pm i$ leaves the whole term still as small as $\epsilon^2$ and $(\alpha^2 + 1)^2$, at least in the vicinity of the desired solutions.  So, using this approximation, Eq.~(\ref{det}) becomes
\begin{equation}
\label{quad-det}
	(\alpha^2 + 1)^2 \pm ~ i \gamma (\alpha^2 + 1) - \epsilon^2 \simeq 0 ~,
\end{equation}
whose solutions are
\begin{equation}
\label{approx}
	\alpha \simeq \pm ~ i  \left(1 \pm ~ i \frac{\gamma}{4} \pm ~ \frac{1}{2} \sqrt{\epsilon^2 - (\frac{\gamma}{2})^2} ~ \right) ~.
\end{equation}

If the three $\pm$'s were chosen independently, there would appear to be eight solutions.  However, the coefficient $\pm i$ of ${\gamma \over 4}$ is an approximation to $\alpha$.  Hence, that $\pm i $ is $+ i $ when $\alpha \approx + i$.  That selects the four actual solutions out the apparent eight, which can also be identified by their having Re$[\alpha] < 0$ for $\gamma > 0$.  So there are really only two $\pm$'s coming from the sequential solution of two quadratic equations, and there is no ambiguity in the term $({\gamma \over 2})^2$.

The separate limits $\epsilon \to 0$ and $\gamma \to 0$ recover the previously understood behaviors of single damped oscillators and coupled, undamped oscillators.

There are evidently three qualitatively different regions, even with both $\gamma \ll 1$ and $\epsilon \ll 1$.  With $\epsilon > \gamma / 2$, the square root term contributes to the oscillation frequency; there are two oscillation frequencies but only one decay rate, which is independent of $\epsilon$.  With $\epsilon < \gamma / 2$, the square root term affects the two decay rates, and there is no splitting of the oscillation frequency degeneracy.  And for $\epsilon \approx \gamma / 2$, the frequencies and decay rates of the two eigenmodes are nearly equal.

In the Appendix, evaluations of Eq.~(\ref{approx}) for three numerical pairs of $\epsilon$ and $\gamma$, representative of the three regions, are compared to the exact values that come from solving Eq.~(\ref{det}).
 
\section{Solution: Eigenvectors}

Let the components of the four eigenvectors be $a$ and $b$:
\begin{equation}
\nonumber
	\mathbf{x_o}
	=
\begin{bmatrix}
	a\\
	b\\
\end{bmatrix}.
\end{equation}
(The eigenvalues $\alpha$ and eigenvectors $\mathbf x_o$, with their components $a$ and $b$, have a four-valued index $i$ to tell which goes with which.  That index $i$ is suppressed when that improves clarity.) 

The lower component of Eq.~(\ref{eq-mot}) tells us that
\begin{equation}
	\label{eigenvec}
	\frac{b}{a} = \frac{- \epsilon}{\alpha^2 +1} \simeq \frac{-\epsilon}{\pm i \gamma / 2 \pm \sqrt{ \epsilon^2 - (\gamma / 2)^2}}~.
\end{equation}
This specifies the four eigenvectors, one for each $\alpha$.  (Recall that $\alpha^2 + 1 = 0$ only for $\epsilon = 0$ and $\alpha^2 + 1 = \pm \epsilon$ for $\gamma = 0$.)  The first expression for $b/a$  with the = sign is exact relative to the initial statement of the problem, i.e., Eq.~(\ref{eq-mot}); the approximate solutions to Eq.~(\ref{quad-det}) are used for the $\simeq$ expression.  The ratio $b/a$ is of order 1 (because $|\alpha^2 + 1| \ll 1$).

Also, $b/a$ is complex.  The phase of each $b/a$ means that in the oscillatory part of the motion corresponding to a single eigenvalue, there is a fixed, non-zero phase difference between the $x_1(t)$ and the $x_2(t)$.  In the language of the plucked string: in the transverse plane, the eigen-motions are elliptical rather than linear (which they would be in the absence of damping).

\section{Real solutions and non-orthogonality}

It is worth returning to the original physical problem and constructing the basis of real eigenfunctions.

The four eigenvalues $\alpha$ are two pairs of complex conjugates.  Label them as $\alpha_{\pm1}$ and $\alpha_{\pm2}$, where there are two, in general, different, negative real parts, each with a pair of conjugate imaginary parts.  The $\alpha_{\pm1}$'s can be assembled into two real eigenfunctions:
\begin{equation}
\nonumber
	\mathbf y_{+1}(t) =
	\text{e}^{\alpha_{+1} t}
	\begin{bmatrix}
	a_{+1} \\
	b_{+1} \\
	\end{bmatrix} +
	\text{e}^{\alpha^*_{+1}t}
	\begin{bmatrix}
	a^*_{+1} \\
	b^*_{+1} \\
	\end{bmatrix}
\end{equation}
\begin{eqnarray}
\nonumber
	\mathbf y_{-1}(t)  & = & i ~ \left(
	\text{e}^{\alpha_{+1}t}
	\begin{bmatrix}
	a_{+1} \\
	b_{+1} \\
	\end{bmatrix} -
	\text{e}^{\alpha^*_{+1}t}
	\begin{bmatrix}
	a^*_{+1} \\
	b^*_{+1} \\
	\end{bmatrix}
	\right)  \nonumber \\
	\nonumber
	& = &  
	\text{e}^{\alpha_{+1}t + i \pi/2}
	\begin{bmatrix}
	a_{+1} \\
	b_{+1} \\
	\end{bmatrix} +
	\text{e}^{\alpha^*_{+1}t + i^* \pi/2}
	\begin{bmatrix}
	a^*_{+1} \\
	b^*_{+1} \\
	\end{bmatrix}~.
\end{eqnarray}
These forms use the facts that $\alpha_{-1} = \alpha_{+1}^*$ and $b_{-1}/a_{-1} = b_{+1}^*/a_{+1}^*$ (true in the original, exact formulation).  Likewise, there are two real functions $\mathbf y_{\pm2}(t)$ similarly constructed out of the conjugate pair $\alpha_{\pm2}$.  Appropriate superpostion of the four $\mathbf y(t)$'s can match the two initial positions and velocities.

The normal modes of linearly coupled, undamped oscillators behave themselves, essentially, like a set of uncoupled oscillators.  Whatever superposition is determined by the initial conditions remains in force for all time.  Many quantities of central importance, such as the kinetic energy, the potential energy, and the total energy of the system are, at any time, just the sum of the contributions from the normal modes.  Since these particular quantities are quadratic in the dynamical variables,  the reduction to a sum over modes requires that cross terms between the contributions of different modes vanish.  And that is the sense in which the normal modes are normal to each other.  However, for the generic case of coupled, damped oscillators, such cross terms are non-zero.  Hence, there are no normal modes --- in spite of there being a complete set of solutions corresponding to the time-dependence eigenvalues.

As long as $\epsilon \ne 0$, for convenience we can choose all four $a_{\pm1,2} = 1$.  Then, for example, at $t=0$:
\begin{eqnarray}
\nonumber
\mathbf y_{+1}(0) \cdot \mathbf y_{+2}(0)  & = & 4 (1 + \text{Re}[b_{+1}]\text{Re}[b_{+2}])  \nonumber \\
\nonumber
 & \ne & 0 \quad \textrm{for}\ \gamma / \epsilon \ne 0 ~.
\end{eqnarray}

Once again, this result is not particularly surprising, except possibly to those imbued with an overwhelming respect for normal modes.  If the coupled, damped system were exactly describable by normal modes, then the total energy would decay steadily as the sum of one or two exponentials.  However, if the corresponding undamped system could exhibit beats, then with the addition of very weak dissipation to only one of the (pre-coupled) degrees of freedom, dissipation should likewise come and go at the beat frequency.  And that is, indeed, one of the possible generic behaviors.

\begin{figure}[h!]
\centering
\includegraphics[width=4.5in]{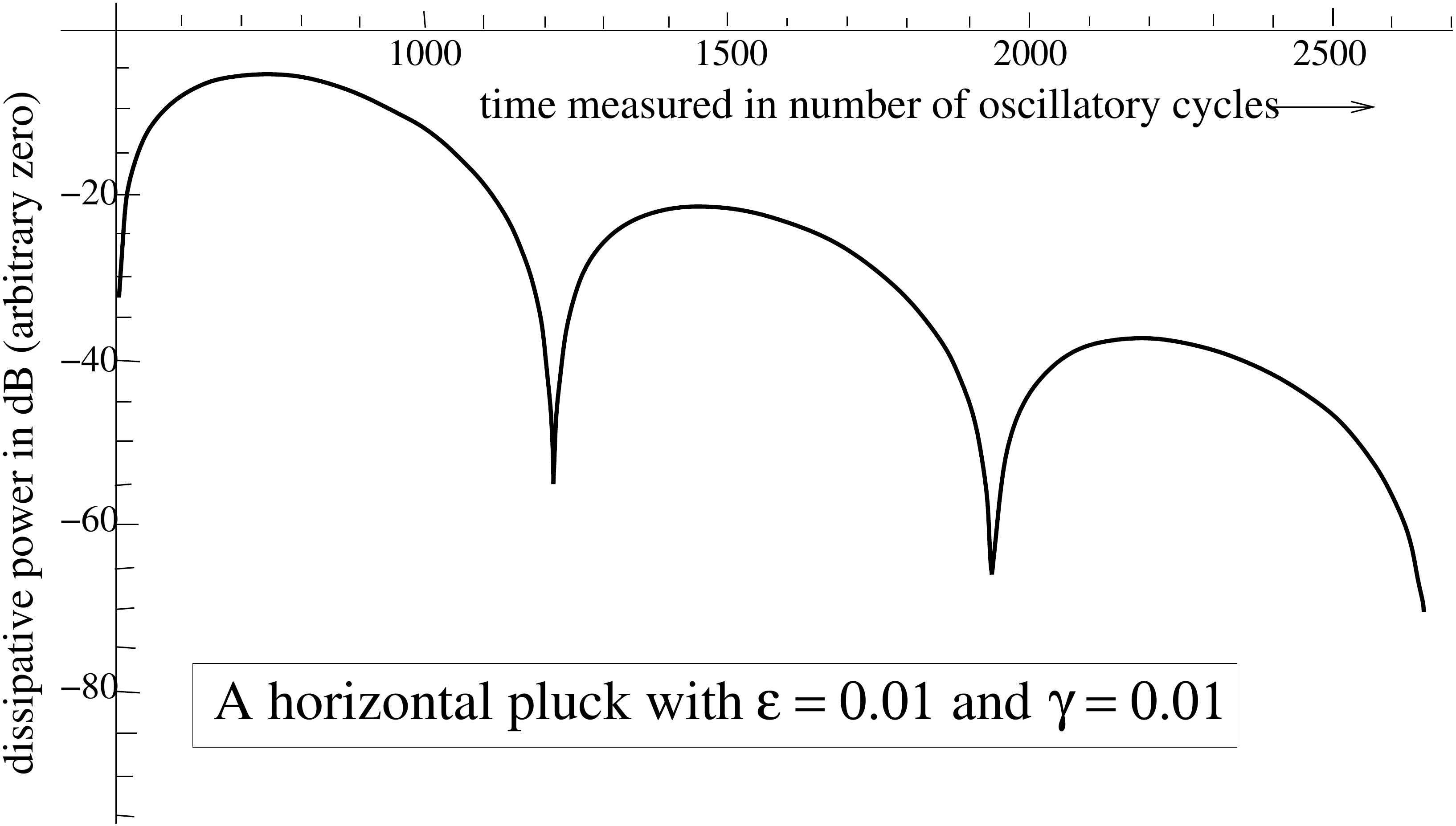}
\caption{The calculated ``sound" of a particular horizontal pluck}
\label{pluck}
\end{figure}

If the simple model analyzed thus far is interpreted as describing a mode doublet of a plucked string, then the produced sound volume would be proportional to the rate of energy lost to damping of the vertical motion of the particular string harmonic.  The instantaneous value of this lost power is $\mathcal P_\text{inst} = \gamma \dot x_2(t)^2$.  In Fig.~\ref{pluck}, the log of the several-cycle-averaged $ \dot x_2(t)^2$ is plotted {\it verus} time for $\epsilon = 0.01$ and $\gamma = 0.01$ with the $t=0$ condition that the pluck is purely in the horizontal (undamped) direction.  The horizontal motion $x_2(t)$ is the lower component of $\mathbf x(t) \equiv \mathbf y_{+1}(t) - \mathbf y_{+2}(t)$.

The numerical parameter values in the Appendix were chosen for convenience of computer entry, with the consideration that they be small but realistic for stringed instruments.  Of these three pairs, the values used for Fig.~\ref{pluck} are the ones that exhibit beats, i.e., there are two distinct oscillation frequencies with the beat period being distinctly shorter than the damping time.  And the horizontal pluck was chosen for display because time must elapse after the pluck before the coupling to the dissipative vertical motion is substantial.  Hence, the dissipated power {\it grows} immediately after the pluck before it subsequently decays.

With a vertical pluck, the dissipation would be evident from the start but would also exhibit the beats between the two eigenfrequencies.  The dissipated power for the other domains of $\epsilon$-$\gamma$ space would look like single or double exponential decay without beats.  Strictly speaking, the single exponential behavior arises at a point in the parameter space.  The two exponential rates become closer and closer as one approaches that point.  On the other side of the point, with only a single exponential, the frequency splitting begins at zero and then grows.

\section{Generic coupling matrix {\bf K}}

The most general possible form of {\bf K} that is close to the identity is
\begin{equation}
\nonumber
	\mathbf{K}
	=
	\begin{pmatrix}
	1+\delta & ~ ~ \eta \\
	\eta ^* & ~ ~ 1-\delta
	\end{pmatrix} ~,
\end{equation}
where $\delta$ is a small detuning of the vertical and horizontal modes and $\eta$ is a small, (in general) complex coupling.  A non-zero phase or imaginary part of $\eta$ would give the undamped normal modes a phase shift between the vertical and horizontal components. If the system under discussion were, indeed, a string, then that phase shift would give an elliptical motion of the undamped normal modes the plane transverse to the string.

When this general {\bf K} is used in Eq.~(\ref{plug-in}), Eq.~(\ref{quad-det}) is replaced by
\begin{equation}
\nonumber
	(\alpha^2 + 1)^2 \pm ~ i \gamma (\alpha^2 + 1) - \delta^2 - \eta \eta ^* \pm ~ i \delta \gamma \simeq 0 ~,
\end{equation}
whose solutions are
\begin{equation}
\nonumber
	\alpha \simeq \pm ~ i  \left(1 \pm ~ i \frac{\gamma}{4} \pm ~ \frac{1}{2} \sqrt{\delta^2 + \eta \eta ^* \pm ~ i \delta \gamma - (\frac{\gamma}{2})^2} ~ \right) ~,
\end{equation}
where the coefficients of $\gamma$ are chosen to have the same sign of $i$ as the $\pm i$ coefficient of the leading 1 term (because their coefficient $i$ is an approximation to the same $\alpha$).  This coincides with the physical condition Re[$\alpha$] $<$ 0 for $\gamma > 0$.

If $\delta$ is negligible, the phase of $\eta$ would only contribute to the phase difference between the components of the eigenvectors, i.e., the analog of Eq.~(\ref{eigenvec}).  This is in addition to the phase difference implicit in Eq.~(\ref{eigenvec}) that arises from the simultaneous real coupling and damping.  The eigenvalues depend only on $\eta \eta^*$ and would be qualitatively as before (with real $\epsilon$).  

If $\eta \eta^*$ were negligible, it would be the simple case of proportional damping.  The oscillation frequencies would be $1 \pm \delta /2$, and only the $1 + \delta /2$ mode would decay, with decay constant $\gamma /2$.

If both $\eta \eta^*$ and $i \delta \gamma$ are relevant, then the square root in the above expression for $\alpha$ is generically complex with a $\pm$ coefficient.  This means that the most general behavior is that the two non-orthogonal modes have different frequencies and different decay rates.  Besides the trivial comment that the magnitudes of the frequency and decay splitting depend on the parameters, it is worth remembering that in practice  the amount of time one has to observe the beats and the second decay rate is limited by the decay. 

\section{Real plucked string sounds}

The subject of what happens in real stringed instruments is a rich one.  The simplest features were recognized in antiquity.  Details have been the subject of serious research over the past 150 years and continues to this day.  Professional journals in which this research is reported include (but are not restricted to) the Journal of the Acoustical Society of America, Acta Acoustica united with Acoustica, and the Journal of Sound and Vibration.  The present discussion has focused on effects that are small compared to the basic oscillatory motion (here normalized to a frequency equal to 1).  Furthermore, attention was restricted to effects that are linear in terms of the differential equations.  This makes the mathematical exercise of relevance to a great number of other physical problems.  However,  in terms of the behavior of an actual plucked string, there may be non-linear effects that are small enough to ignore relative to that 1 but are significant in the present context.  Many effects have been discussed in the scientific literature.  But there is no consensus as to which is ``the correct one."  In fact, the relative importance of different mechanisms may depend on the kind of instrument and even the individual instrument.  The most basic lesson to take away is that, when dealing with the interactions of frequency degenerate systems, very tiny forces can lead to clearly visible effects if they can act over many cycles of the primary oscillation.

The following are just a few of possible contributions to the parameters in {\bf K}.  One contribution to the ``detuning" $\delta$ might be an approximate description of consequences of the up-down motion of the bridge end of the string being much larger than the horizontal motion.  For a particular vertical mode, the bridge end is no longer an actual node, and the effective length of the string is longer than were it a node.  A small skewness of the elastic constants relative to vertical/horizontal (i.e., $\epsilon$ or Re[$\eta$]) can arise from imperfections and asymmetries in the materials of the string and its supports, which can rotate the principle axes of the restoring forces relative to exactly parallel and perpendicular to the head.  The longitudinal stretching of the string is usually successfully ignored at the level of the 1, but its small effects can be relevant when one looks closer.  Sometimes an effort is made to approximate those effects and represent them as contributions to the parameters of the generic linear system.  Players of stringed instruments easily notice that the motion in the transverse plane is typically elliptical, and, furthermore, those ellipses typically precess.  This would happen even if the only deviation from degeneracy were a detuning of the two components and the initial pluck was not exactly in one of the normal mode directions.  A separate issue is whether the undamped normal modes themselves describe elliptical motion, i.e., whether $\eta$ has a phase.  This would not arise from a combination of linear springs, but it cannot be ruled out as an effective, approximate description of the combined effects of linear and non-linear dynamics.

\begin{figure}[h!]
\centering
\includegraphics[width=2.7in]{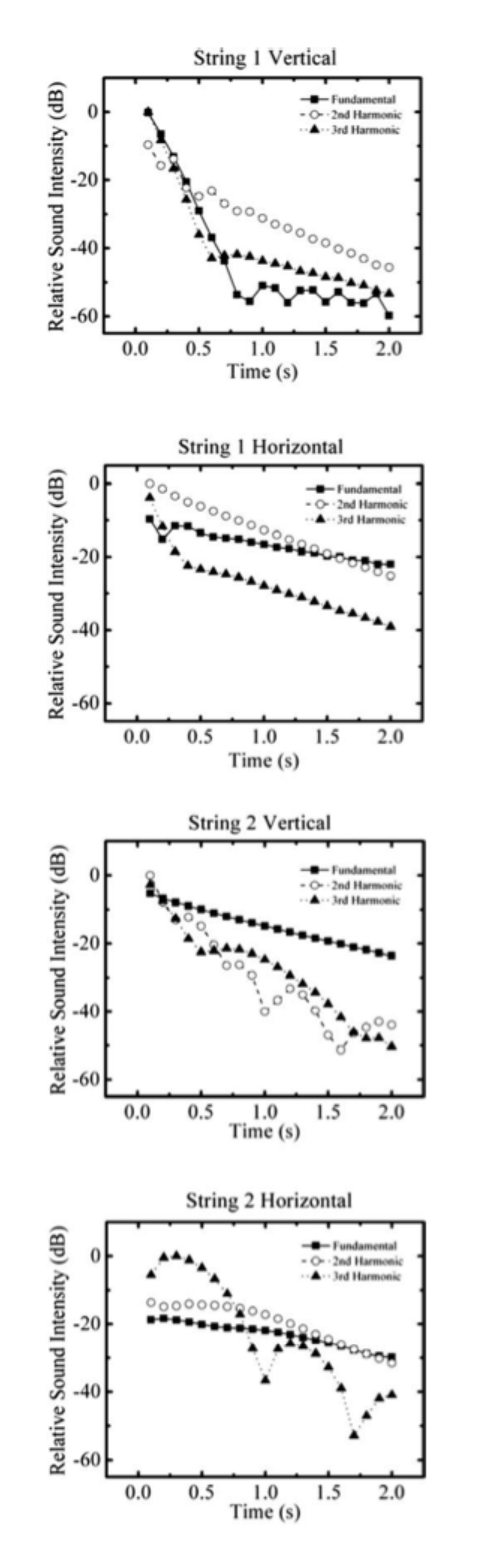}
\caption{Measured loudness in dB (log scale) {\it vs.} time for string harmonics, copied from Ref.~\onlinecite{moore}}
\label{plucked-strings}
\end{figure}

Even given the uncertainties described above, it is interesting to look at a case of measured, real string motion where the investigators projected out the time evolution of particular, individual modes. 
Measurements of the sounds of plucked banjo strings were published by Moore and Stephey.\cite{moore}   For one aspect of their experimental survey, they damped all strings but the first, plucked it, first in the vertical direction (perpendicular to the banjo head) and then in the horizontal.  They did the same for the second string.  They recorded the sounds and analyzed them into Fourier components.  In Fig.~\ref{plucked-strings} (copied from Fig.~5 of Ref. \onlinecite{moore}) the (logarithmic) sound intensities for the first three harmonics of each plucked string are displayed as functions of time.  
Each harmonic acts as a separate nearly degenerate, coupled, damped pair.  As those authors noted in their paper, evident are single exponential decays, double exponential decays, and decays with prominent beats modulating a single overall decay rate.

The qualitative agreement with the forms of the time dependence predicted by the initial model of {\bf K} with a single, real parameter $\epsilon$ does not rule out the more general form.  Rather, it suggests that the region of four dimensional parameter space of the generic linear model populated by the experimentally observed modes is roughly two dimensional and might be well-approximated by an effective $\epsilon$.  More measurements would be needed to do better.
  
With just the naked eye, some of this behavior is typically visible on a stringed instrument.  In particular, there is usually at least one string that after a pluck exhibits beats.  Instead of decreasing steadily, its amplitude gets smaller and larger again a couple or even several times before it dies completely.  (Generally, each maximum is smaller than the previous one.)  Identifying double exponential decay is harder, just judging by eye.  However, there are typically some pluck responses with a dramatic initial fall off and a surprisingly long tail.

This phenomenon is actually a very important aspect of banjo sound.  The banjo is an instrument where the degeneracy is often four- and even six-fold, not just the two of a single string.  That is because, as normally tuned and played, the undamped strings are often in unison or share harmonics (e.g., the second harmonic of one string is degenerate with the third harmonic of another).  And the design of the bridge facilitates coupling between between all of the strings.

\appendix
\section{Exact/approximate eigenvalue comparison}
Computer math packages include the exact solution for the roots of the general quartic polynomial.  Hence, their numerical evaluations of Eq.~(\ref{det}) are unassailable.  The table below gives comparison of the approximate eigenvalues from Eq.~(\ref{quad-det}) to the exact eigenvalues from Eq.~(\ref{det}) for $\epsilon = 0.01$ and $\gamma = 0.01$, 0.02, and 0.03.

\begin{table}[h!]
\centering
\caption{Eigenvalue exact/approximate comparison}
\begin{ruledtabular}
\begin{tabular}{l c c c c}
$\epsilon$ & $\gamma$ & method & Re[$\alpha$] & $i$Im[$\alpha$] \\
\hline
0.01 & 0.01 & exact & -0.002504 & $\pm i$ 0.995679 \\
 & & & -0.002496 & $\pm i$ 1.00431 \\
0.01 & 0.01 & approx & -0.0025 & $\pm i$ 0.995670 \\
 & & & -0.0025 & $\pm i$ 1.00433 \\
\hline
0.01 & 0.02 & exact & -0.005352 & $\pm i$ 0.9996 \\
 & & & -0.004648 & $\pm i$ 1.0003 \\
0.01 & 0.02 & approx & -0.005 & $\pm i$  \\
 & & & -0.005 & $\pm i$  \\
\hline
 0.01 & 0.03 & exact & -0.01309021 & $\pm i$ 0.999856 \\
 & & & -0.00190978 & $\pm i$ 1.00001 \\
0.01 & 0.03 & approx & -0.01309017 & $\pm i$  \\
 & & & -0.00190983 & $\pm i$  \\
\end{tabular}
\end{ruledtabular}
\end{table}


\newpage

\begin{thebibliography}{99}

\bibitem{MGM} M. Gell-Mann and A. Pais, ``Behavior of neutral particles under charge conjugation,"Phys. Rev. \textbf{97} (5) 1387--1389 (1955).

\bibitem{CP} J. H. Christenson, J. W. Cronin, V. L. Fitch, and R. Turlay, ``Evidence for the 2$\pi$ decay of the $K_2^{~o}$ meson," Phys. Rev. Lett. \textbf{13} (4) 138--140 (1964).

\bibitem{kabir}  P. K. Kabir, \textit{The CP Puzzle}, Academic,
New York (1968).

\bibitem{caughey} T. K. Caughey, ``Classical Normal Modes in Damped Linear Dynamic Systems," J. Appl. Mech. \textbf{27E} 269-271 (1960).


\bibitem{caughey2} T. K. Caughey and M. E. J. O'Kelly, ``Classical normal modes in damped linear dynamic systems," J. Appl. Mech. \textbf{32} (3), 583-588 (1965).

\bibitem{weinreich} Gabriel Weinreich, ``Coupled piano strings,"J. Acoust. Soc. Am. \textbf{62} (6) 1474--1484 (1977).

\bibitem{schmidt} P. J. Schmid and D. S. Henningson,
{\it Stability and Transition in Shear Flows}, Springer-Verlag, New York, 2001.

\bibitem{tref} L. N. Trefethen and Mark Embree, {\it Spectra and Pseudospectra -- The Behaviour of Nonnormal Matrices and Operators,} Princeton University Press, Princeton and Oxford, 2005.

\bibitem{winstein} A. Alavi-Harati, {\it et al.} (the KTeV collaboration), ``Measurements of the Decay $K_L \rightarrow e^+ e^- \mu^+ \mu^-$," Phys. Rev. Lett. \textbf{90} (14) 1801-1805 (2003).

\bibitem{quant} Jonathan L. Rosner and Scott A. Slezak, ``Classical illustrations of CP violation in kaon decays," Am. J. Phys. {\bf 60}  (1) 44-49 (2001).

\bibitem{springs} Frank Karioris and Kenneth Mendelson, ``A novel coupled oscillation demonstration," Am. J. Phys. {\bf 60}  (6) 508-512 (1992).

\bibitem{pendulums} M. Maianti, S. Pagliara, G. Galimberti, and F. Parmigiani, 
``Mechanics of two pendulums coupled by a stressed spring," Am. J. Phys. {\bf 77}  (9) 834-838 (2014).

\bibitem{wilbur} Matthew Mewes, ``The Slinky Wilberforce pendulum: A simple coupled oscillator,"  Am. J. Phys. {\bf 82}  (3) 254-256 (2014).

\bibitem{rahul} Rahul Bale and Rama Govindarajan, ``Transient growth and why we should care about it," Resonance \textbf{15} (5)  441--457 (2010); this article offers a fine, elementary introduction to the subjects addressed in Ref.s 
\onlinecite{schmidt} and \onlinecite{tref}.

\bibitem{quantum} M. Bhattacharya, M. J. A. Stoutimore, K. D. Osborn and Ari Mizel, ``Understanding the damping of a quantum harmonic oscillator coupled to a two-level system using analogies to classical friction," Am. J. Phys. {\bf 80}  (9) 810-815 (2012).


\bibitem{weinreich2} The analysis of coupled, damped strings in Ref.~\onlinecite{weinreich} likewise uses near degeneracy as a key approximation.  However, the methods and tools of analysis used there assume considerably greater sophistication on the part of the reader.  Also, with somewhat different goals, several of the very general features are not remarked upon as such.

\bibitem{moore} Laurie A. Stephey and Thomas R. Moore, ``Experimental investigation of an American five-string banjo," J. Acoust. Soc. Am. \textbf{124} (5) 3276--3283 (2008).
\end{thebibliography}
\end{document}